\documentclass[aps,prd,showpacs,preprintnumbers,superscriptaddress,nofootinbib,twocolumn]{revtex4}
\usepackage[dvips]{graphicx}
\usepackage{bm,latexsym,amsmath,amssymb,amsfonts,mathrsfs}
\usepackage{color}
\input{colordvi.tex}
\usepackage{color}
\newcommand*{\D}{{\rm d}}

\newcommand*{\mpl}{M_{\rm Pl}}
\begin{document}

\title{Cosmological perturbations in a healthy extension of Ho\v{r}ava gravity}

\author{Tsutomu~Kobayashi}
\email[Email: ]{tsutomu"at"resceu.s.u-tokyo.ac.jp}
\affiliation{Department of Physics, Waseda University, Okubo 3-4-1,
Shinjuku, Tokyo 169-8555, Japan \footnote{Present address: Research
Center for the Early Universe (RESCEU), Graduate School of Science, The
University of Tokyo, Tokyo 113-0033, Japan }}
\author{Yuko~Urakawa}
\email[Email: ]{yuko"at"gravity.phys.waseda.ac.jp}
\affiliation{Department of Physics, Waseda University, Okubo 3-4-1, Shinjuku, Tokyo 169-8555, Japan}
\author{Masahide~Yamaguchi}
\email[Email: ]{gucci"at"phys.titech.ac.jp}
\affiliation{Department of Physics and Mathematics, Aoyama Gakuin
University, Sagamihara 229-8558, Japan \footnote{Present address:
Department of Physics, Tokyo Institute of Technology, Tokyo 152-8551,
Japan}}
  
\begin{abstract}
In Ho\v{r}ava's theory of gravity, Lorentz symmetry is broken in
exchange for renormalizability, but the original theory has been argued
to be plagued with problems associated with a new scalar mode stemming
from the very breaking of Lorentz symmetry. Recently, Blas, Pujol\`{a}s,
and Sibiryakov have proposed a healthy extension of Ho\v{r}ava
gravity, in which the behavior of the scalar mode is improved.  In
this paper, we study scalar modes of cosmological perturbations in
extended Ho\v{r}ava gravity.  The evolution of metric and density
perturbations is addressed analytically and numerically. It is shown
that for vanishing non-adiabatic pressure of matter the large scale
evolution of cosmological perturbations converges to that
described by a single constant, $\zeta$, which is an analog of a
curvature perturbation on the uniform-density slicing commonly used in
usual gravitational theories. The subsequent evolution is thus
determined completely by the value of $\zeta$.
\end{abstract}

\preprint{WU-AP/306/10}
\maketitle

\section{Introduction}

Recently, Ho\v{r}ava has proposed a new theory of quantum
gravity~\cite{Horava}.  The basic idea of the theory is abandoning the
Lorentz invariance, with which the theory is made power-counting
renormalizable.  The broken Lorentz invariance results in a preferred
foliation of spacetime by three-dimensional spacelike hypersurfaces, and
thus allows to add higher spatial curvature terms to the gravitational
Lagrangian as well as to modify the kinetic term of
the graviton.  Obviously, some notable features of Ho\v{r}ava gravity
become manifest at high energies, which motivates studying the early
universe based on Ho\v{r}ava's theory~\cite{Calc, Kiritsis,
earlycosmology}. The effect of the broken Lorentz invariance, however,
persists down to low energies.  This not only brings interesting
consequences regarding in particular dark matter in the
universe~\cite{DM_M}, but also causes potential problems stemming from
an additional scalar degree of freedom of the graviton that inevitably
appears as a result of the reduced symmetry of the theory~\cite{Charm,
Li, BPS_1, K-A, Henn}.  Some of the troubles are cured in the
projectable version of Ho\v{r}ava gravity~\cite{DM_M, Caustic} and some
are still controversial.  Several problems have also been reported at a
phenomenological level, including a large isocurvature
perturbation~\cite{KUY} and the absence of static stars~\cite{I-M} (see
also~\cite{Greenwald}).

It is argued by Blas, Pujol\`{a}s, and Sibiryakov that original
Ho\v{r}ava gravity can be extended in such a way that the lapse function
$N$ may depend on the spatial coordinate (i.e., the theory is not
projectable) and terms constructed from a 3-vector $\partial_i\ln N$ are
included in the action~\cite{BPS}, which makes $N$ dynamical~\cite{LLMU}. With
the appropriate choice of the new terms, the resultant theory will be
free from the problems and pathologies reported in the
literature~\cite{Charm, Li, BPS_1, K-A, Henn}.  The extended theory
could still suffer from strong coupling at low energies~\cite{Papa}, but
it can be evaded by taking into account higher spatial derivative terms
in the action~\cite{BPS:comment}. See also the recent paper~\cite{KP} for
the strong-coupling issue in the extended version of Ho\v{r}ava gravity.

So far cosmological perturbations have been explored in projectable and
non-projectable versions of Ho\v{r}ava gravity in the vast
literature~\cite{KUY, Gao, Zatta, W-M, WWM, GKS}.  The aim of this paper
is to go on in this direction and to study cosmological
perturbations in healthfully extended Ho\v{r}ava gravity of~\cite{BPS}.
We begin with presenting the most general equations for scalar
perturbations on a flat cosmological background in extended Ho\v{r}ava
gravity, and then solve them both analytically and numerically.  As for
the other aspects of extended Ho\v{r}ava gravity, spherically symmetric
solutions have been worked out in~\cite{Kiri-Sph}.

The paper is organized as follows.  In the next section we review how
non-projectable Ho\v{r}ava gravity can be healthfully extended,
following~\cite{BPS}.  Then, in Sec.~III, the background evolution and
perturbation equations are given within the framework of extended
Ho\v{r}ava gravity.  In Sec.~IV the perturbation equations are solved
analytically and numerically.  We draw our conclusions in Sec.~V.

\section{A healthy extension of Ho\v{r}ava gravity}

The constituent variables in Ho\v{r}ava gravity are
$N$, $N_i$, and $g_{ij}$, which correspond to the lapse function, the shift vector, and
the spatial metric, respectively, in the $(3+1)$ decomposition of a spacetime.
The non-projectable version of Ho\v{r}ava gravity
is allowed to include terms constructed from the 3-vector,
\begin{eqnarray}
a_i:=\partial_i\ln N(t,\Vec{x}),
\end{eqnarray}
in the Lagrangian. Along with the line of Refs.~\cite{BPS, SVW},
we consider the extended Ho\v{r}ava gravity described by
\begin{eqnarray}
S&=&\frac{\mpl^2}{2}\int \D t\D^3x \sqrt{g}N\left({\cal L}_K-{\cal V}[g_{ij}, a_i]\right)
\nonumber\cr&&\quad
+\int\D t\D^3x\sqrt{g}N{\cal L}_{\rm m}.
\end{eqnarray}
The ``kinetic term'' is given by
\begin{eqnarray}
{\cal L}_K& =&K_{ij}K^{ij}-\lambda K^2,\label{kinetic_term}
\end{eqnarray}
with the extrinsic curvature defined by
$K_{ij}=\left(\dot{g}_{ij}-\nabla_iN_j-\nabla_jN_i\right)/2N$,
while the ``potential'' is given by ${\cal V}[g_{ij}, a_i]=\sum_{z=1}^{3}{\cal V}_z$, where
\begin{eqnarray}
{\cal V}_1&=&-R-\eta a_ia^i,
\\
{\cal V}_2&=&\mpl^{-2}(
g_2 R^2+g_3R_{ij}R^{ij}
\nonumber\\&&\quad
+\eta_2 a_i\Delta a^i+\eta_3 R\nabla_ia^i+...
),
\\
{\cal V}_3&=&\mpl^{-4}(
g_4 R\Delta R+g_5\nabla_iR_{jk}\nabla^iR^{jk}
\nonumber\\&&\quad
+\eta_4a_i\Delta^2a^i+\eta_5\Delta R\nabla_ia^i+...).
\end{eqnarray}
Here, the Ricci tensor $R_{ij}$ and Ricci scalar $R$ are constructed from $g_{ij}$,
$\nabla_i$ is the covariant derivative with respect to $g_{ij}$, $\Delta:=\nabla_i\nabla^i$,
and $\dot{}:=\partial_t$.
In ${\cal V}_2$ and ${\cal V}_3$
we only write explicitly the terms that will be relevant to
{\em scalar-type, linear perturbations} on a {\em flat} cosmological background.
For instance, terms such as $a_ia_jR^{ij}$ and $(a_ia^i)^2$ are not relevant to
linear perturbations on a flat cosmological background.
A parity-violating term $\epsilon^{ijk}R_{il}\nabla_jR^{l}_{\;k}$,
which appears in the original version of Ho\v{r}ava gravity with the detailed valance condition~\cite{Horava},
is not relevant to scalar-type perturbations.
The potential presented above is therefore the most general one
in the situation we are considering.
${\cal L}_{\rm m}$ is the Lagrangian of matter.
In this paper we simply assume that matter is minimally coupled to gravity.

The symmetry of the theory is invariance under the foliation-preserving diffeomorphism transformations:
$t\to \tilde t(t),\; x^i\to\tilde x^i(t, \Vec{x})$.
Under the infinitesimal transformation,
$t\to t+f(t)$, $x^i\to x^i+\xi^i(t,\Vec{x})$,
the variables transform as
\begin{eqnarray}
N&\to& N-f\dot N-\dot fN-\xi^i\partial_iN,
\nonumber\\
N_i&\to& N_i-\nabla_i\xi^jN_j-\xi^j\nabla_jN_i-\dot\xi^j g_{ij}-\dot fN_i-f\dot N_i,
\nonumber\\
g_{ij}&\to& g_{ij}-\dot g_{ij}f-g_{ik}\nabla_j\xi^k-g_{jk}\nabla_i \xi^k.
\label{transformaton}
\end{eqnarray}

As a result of broken general covariance, a new scalar degree of freedom appears
in addition to the usual helicity-2 polarizations of the graviton,
and this scalar graviton could be a ghost depending on the value of $\lambda$.
Since the structure of the kinetic term~(\ref{kinetic_term})
is the same as in the original version of Ho\v{r}ava gravity,
the time kinetic term of the scalar graviton remains the same.
Therefore,
the condition for avoiding ghosts is~\cite{Horava, BPS}
\begin{eqnarray}
\mathscr{A}:=\frac{3\lambda-1}{\lambda-1} >0.\label{ghostfree}
\end{eqnarray}
However, the sound speed squared is negative for $\mathscr{A}>0$ in original Ho\v{r}ava gravity,
indicating that the scalar graviton is unstable~\cite{W-M, K-A}.
This fact itself does not necessarily mean that the theory is ``unhealthy,''
because whether or not an instability really causes a trouble depends upon its time scale~\cite{I-M}.
In the above ``healthy'' extension of Ho\v{r}ava gravity,
such an instability can be made absent from the beginning for $\mathscr{A}>0$
by the appropriate choice of the potential terms of $a_i$~\cite{BPS}.

The Hamiltonian constraint is derived by varying the action with respect to $N$:
\begin{eqnarray}
{\cal L}_K+{\cal V}+\delta{\cal V}+\frac{2}{\mpl^2}\rho=0,
\end{eqnarray}
where
\begin{eqnarray}
\delta{\cal V}&:=&2\eta\nabla_i a^i
-\frac{2\eta_2}{\mpl^2}\Delta \nabla_ia^i
+\frac{\eta_3}{\mpl^2}\Delta R
\nonumber\\&&-\frac{2\eta_4}{\mpl^4}\Delta^2\nabla_ia^i+\frac{\eta_5}{\mpl^4}\Delta^2R+...\,.
\end{eqnarray}
We have written here explicitly the terms that are relevant at
zeroth and linear order in our cosmological setting.  The matter energy
density is defined as $ \rho :=-{\cal L}_{\rm m}-N \delta{\cal L}_{\rm
m}/\delta N.  $ Variation with respect to the shift vector gives the
momentum constraint equations:
\begin{eqnarray}
\nabla_j\pi^{ij}=\frac{1}{\mpl^2}J^i,
\end{eqnarray}
where $\pi^{ij}:=K^{ij}-\lambda Kg^{ij}$ and $J^i:=-N\delta{\cal L}_{\rm m}/\delta N_i$.
Finally, the evolution equations are derived from variation with respect to $g_{ij}$:
\begin{widetext}
\begin{eqnarray}
&&2\left(K_{ik}K^k_{\;j}-\lambda KK_{ij}\right)-\frac{1}{2}g_{ij}{\cal L}_K
+\frac{1}{N\sqrt{g}}\frac{\partial }{\partial t}\left(\sqrt{g}\pi^{kl}\right)g_{ik}g_{jl}
-\frac{1}{N}\nabla^k\left(\pi_{ij}N_k\right)
+\frac{1}{N}\nabla^k\left(\pi_{ik}N_j\right)
+\frac{1}{N}\nabla^k\left(\pi_{kj}N_i\right)
\nonumber\cr&&\quad
+\frac{1}{N}\Delta Ng_{ij}-\frac{1}{N}\nabla_i\nabla_jN
+R_{ij}-\frac{1}{2}g_{ij}R+
\frac{2g_2}{\mpl^2}\left(\nabla_i\nabla_j-g_{ij}\Delta\right)R
+\frac{g_3}{\mpl^2}\left[\nabla_i\nabla_jR-\Delta\left(R_{ij}+\frac{1}{2}g_{ij}R\right)\right]
\nonumber\cr&&\qquad
+\frac{2g_4}{\mpl^4}\left(\nabla_i\nabla_j-g_{ij}\Delta\right)\Delta R
-\frac{g_5}{\mpl^4}\left[\nabla_i\nabla_j\Delta R-\Delta^2\left(R_{ij}+\frac{1}{2}g_{ij}R\right)\right]
\nonumber\cr&&\quad\qquad
+\left(\nabla_i\nabla_j-g_{ij}\Delta\right)\left(\frac{\eta_3}{\mpl^2}\nabla_ka^k
+\frac{\eta_5}{\mpl^4}\Delta\nabla_ka^k\right)+...
=\frac{T_{ij}}{\mpl^2},
\end{eqnarray}
\end{widetext}
where $T_{ij}:={\cal L}_{\rm m}g_{ij}-2\delta{\cal L}_{\rm m}/\delta g^{ij}$.
Here also we have written explicitly only the terms linear in $R_{ij}$ and $a_i$;
the other possible terms are not relevant in the present paper.

The matter action is invariant under the infinitesimal transformation~(\ref{transformaton}),
which results in
\begin{eqnarray}
&&\int\D^3 x\left[
\frac{\sqrt{g}N}{2}\dot g_{ij}T^{ij}+N\partial_t\left(\sqrt{g}\rho\right)+N_i\partial_t\left(
\sqrt{g}J^i\right)
\right]
\nonumber\\&&\quad
=0,\label{en_cn}
\end{eqnarray}
and
\begin{eqnarray}
&&\nabla^jT_{ij}+a_i\rho+a_jT_i^{\;j}
-\frac{1}{N\sqrt{g}}\partial_t\left(\sqrt{g}J_i\right)-\frac{N_i}{N}\nabla_jJ^j
\nonumber\\&&\qquad\qquad
-\frac{J^j}{N}\left(\nabla_jN_i-\nabla_iN_j\right)=0.
\label{mom_cn}
\end{eqnarray}
Note that the energy conservation equation~(\ref{en_cn})
is given by the integration over the whole space,
as the gauge parameter $f(t)$ is space-independent.

\section{Cosmology of extended Ho\v{r}ava gravity}

\subsection{Background equations}

To describe the background evolution of the universe we write
$N=1$, $N_i =0$, and $g_{ij}=a^2(t)\delta_{ij}$, assuming that the three-dimensional spatial section is flat.
The Hamiltonian constraint reduces to
\begin{eqnarray}
\frac{3(3\lambda-1)}{2}H^2 = \frac{\rho}{\mpl^2},\label{Fr1}
\end{eqnarray}
where $H:=\dot a/a$ and $\rho$ is the background value of the energy density of matter,
while the evolution equation reads
\begin{eqnarray}
-\frac{3\lambda-1}{2}\left(3H^2+2\dot H\right)=\frac{p}{\mpl^2},\label{Fr2}
\end{eqnarray}
where $p$ is the background value of the isotropic pressure of matter, $T_i^{\;j} =p\delta_i^{\;j}$.
From the above equations we may naturally identify
the gravitational constant in Friedmann-Robertson-Walker cosmology as
$8\pi G_c:=2\mpl^{-2}(3\lambda-1)^{-1}$.

It follows from Eqs.~(\ref{Fr1}) and~(\ref{Fr2}) that
\begin{eqnarray}
\dot\rho +3H(\rho+p)=0.
\end{eqnarray}
Obviously, the energy conservation equation derived from~(\ref{en_cn}),
\begin{eqnarray}
\int \D^3 x\;a^3\left[\dot\rho+3H\left(\rho+p\right)\right]=0,
\end{eqnarray}
is satisfied. Note that
in contrast to the case of projectable Ho\v{r}ava gravity,
we now have the local Hamiltonian constraint and hence
``dark matter as an integration constant''~\cite{DM_M} does not appear.

We see that nothing special happens at the background level,
except that the gravitational constant $G_c$ differs from
the locally measured value of Newton's constant, $G_N=\mpl^{-2}(1-\eta/2)^{-1}$~\cite{BPS}.

\subsection{Cosmological perturbations}

General scalar perturbations of the lapse function, shift vector,
and the spatial metric
can be written as
\begin{eqnarray}
N=1+\phi(t,\Vec{x}), \quad N_i=a^2\partial_i\beta(t,\Vec{x}),\nonumber\\
g_{ij} = a^2\left[1-2\psi(t,\Vec{x})\right]\delta_{ij}+2a^2\partial_i\partial_jE(t,\Vec{x}).
\end{eqnarray}
Note that in the cosmological background of Ho\v{r}ava gravity we are considering,
linear perturbations can be decomposed into scalar, vector, and tensor perturbations
because all the background quantities are homogeneous and isotropic.
Under a scalar gauge transformation, $t\to t+f(t)$, $x^i\to x^i+\partial^i\xi(t,\Vec{x})$,
these perturbations transform as
\begin{eqnarray}
\phi\to\phi-\dot f,\;\;\beta\to\beta-\dot\xi,\nonumber\\
\psi\to \psi+H f,\;\;E\to E-\xi.
\end{eqnarray}
By using the spatial gauge transformation we may set $E=0$.
Note, however, that the temporal gauge degree of freedom
does not help to remove $\phi$ or $\psi$ because $f$ is a function of $t$ only.

The linearized Hamiltonian constraint is given by
\begin{eqnarray}
&&3(3\lambda-1)H \left(\dot\psi+H\phi +\frac{1}{3}\nabla^2\beta\right)
\nonumber\\&&\quad
-2\frac{\nabla^2}{a^2}\psi+\eta\frac{\nabla^2}{a^2}\phi+\frac{\nabla^2}{a^2}\delta{\cal H}
+\frac{\delta\rho}{\mpl^2} =0,\label{p-ham}
\end{eqnarray}
where
\begin{eqnarray}
\delta{\cal H}(t, \Vec{x})&:=&
-\frac{\eta_2}{\mpl^2}\frac{\nabla^2}{a^2}\phi+\frac{2\eta_3}{\mpl^2}\frac{\nabla^2}{a^2}\psi
\nonumber\\&&\quad
-\frac{\eta_4}{\mpl^4}\frac{\nabla^4}{a^4}\phi+\frac{2\eta_5}{\mpl^4}\frac{\nabla^4}{a^4}\psi.
\end{eqnarray}
The momentum constraints at linear order read
\begin{eqnarray}
(3\lambda-1)\left(\dot\psi+H\phi\right)+(\lambda-1)\nabla^2\beta
=\frac{\delta J}{\mpl^2},\label{p-m-c}
\end{eqnarray}
where we write $J_i=\partial_i\delta J$.
The evolution equations can be written as
\begin{eqnarray}
{\cal G}_{\rm T}\delta_i^{\;j}-\left(\frac{\partial_i\partial^j}{\nabla^2}
-\frac{1}{3}\delta_i^{\;j}\right){\cal G}_{\rm TL}
=\frac{\delta T_i^{\;j}}{\mpl^2},
\end{eqnarray}
where
\begin{eqnarray}
{\cal G}_{\rm T}&=&(3\lambda-1)\left[\ddot\psi+3H\dot\psi+H\dot\phi+\left(3H^2+2\dot H\right)\phi
\right]
\nonumber\\&&
+(\lambda -1)\nabla^2\left(\dot\beta+3H\beta\right)+\frac{2}{3}{\cal G}_{\rm TL}
\end{eqnarray}
and
\begin{eqnarray}
{\cal G}_{\rm TL}&=&\nabla^2\left(\dot\beta+3H\beta\right)+\frac{\nabla^2}{a^2}\left(\phi -\psi\right)
+\frac{\nabla^2}{a^2}\delta{\cal E},
\end{eqnarray}
with
\begin{eqnarray}
\delta{\cal E}(t, \Vec{x})&=&-\frac{8g_2+3g_3}{\mpl^2}\frac{\nabla^2}{a^2}\psi
-\frac{8g_4-3g_5}{\mpl^4}\frac{\nabla^4}{a^4}\psi
\nonumber\\&&\;\; -\frac{\eta_3}{\mpl^2}\frac{\nabla^2}{a^2}\phi-\frac{\eta_5}{\mpl^4}\frac{\nabla^4}{a^4}\phi.
\end{eqnarray}
Neglecting for simplicity the anisotropic stress perturbation, we have
\begin{eqnarray}
{\cal G}_{\rm T}&=&\frac{\delta p}{\mpl^2},\label{p-ev-t}\\
{\cal G}_{\rm TL}&=&0.\label{p-ev-tl}
\end{eqnarray}

The perturbed energy conservation equation reduces to
\begin{eqnarray}
&&\int\D^3 x\;a^3\left[\dot{\delta\rho}+3H\left(\delta\rho+\delta p\right)
-3\dot\psi(\rho+p)
\right]
\nonumber\\&&
+\int\D^3 x\;a^3\left[\dot\rho+3H\left(\rho+p\right)\right](\phi-3\psi)
=0.\label{p-e-c}
\end{eqnarray}
The second line vanishes thanks to the background equation.
Combining the Hamiltonian constraint~(\ref{p-ham}) and
the evolution equations~(\ref{p-ev-t}) and~(\ref{p-ev-tl}), and using the background equations, one finds
\begin{eqnarray}
&&\mpl^{-2}\left[\dot{\delta\rho}+3H(\delta\rho+\delta p)-3\dot\psi(\rho+p)\right]
\nonumber\\
&&\;
=\frac{\nabla^2}{a^2}\left[
2\left(\dot\psi+H\psi\right)-\eta\left(\dot\phi+H\phi\right)
-\left(\dot{\delta{\cal H}}+H\delta{\cal H}\right)
\right]
\nonumber\\
&&\;\;\;
+\nabla^2\left[(1-3\lambda)\dot H \beta
-2H \left(\dot\beta+3H\beta\right)\right].
\label{comb-ein-eq}
\end{eqnarray}
Since the right hand side is a total derivative, the first line of Eq.~(\ref{p-e-c})
also vanishes, and hence Eq.~(\ref{p-e-c}) is automatically satisfied.
Note, however, that the local
equation~(\ref{comb-ein-eq}) is stronger than the integrated equation~(\ref{p-e-c}).
The momentum conservation equation gives
\begin{eqnarray}
\dot{\delta J}+3H\delta J-(\rho+p)\phi-\delta p =0.
\end{eqnarray}
This equation can also be derived using the momentum constraint~(\ref{p-m-c}) and
the evolution equations~(\ref{p-ev-t}) and~(\ref{p-ev-tl}) together with the background equations.
Therefore, the energy and momentum conservation equations do not give rise to any
independent equations.

In order to obtain a closed set of equations,
we need a matter equation of motion.
Suppose that the matter equation of motion takes of the form
\begin{eqnarray}
&&\dot{\delta\rho}+3H(\delta\rho+\delta p)-3\dot\psi(\rho+p)
\nonumber\\&&\quad -\frac{\nabla^2}{a^2}\left[\delta J+a^2(\rho+p)\beta\right]
 = \frac{\nabla^2}{a^2}\delta{\cal F}(t,\Vec{x}),
\end{eqnarray}
where the concrete form of $\delta{\cal F}$ depends on the matter field
one is considering.  A matter field that respects four-dimensional
general covariance is conserved locally, and consequently, $\delta {\cal
F}=0$.  However, in Ho\v{r}ava gravity it is natural to consider that
matter fields respect only the foliation-preserving diffeomorphism, and
hence are not conserved locally in general. For example, a scalar field
Lagrangian presented in~\cite{Calc, Kiritsis, WWM} leads to $\delta
{\cal F}\neq 0$.  In this paper, we assume that $\delta{\cal F}$ arises
only from the ultraviolet (UV) effect and is suppressed by $\Delta
/M_{\rm m}^2$, where $M_{\rm m}$ is a typical mass scale.  $M_{\rm m}$
is not necessarily of the same order of $\mpl$ and may be different for
different matter fields, but here we simply assume that the
scale $M_{\rm m}$ is sufficiently high.  Using the momentum
constraint~(\ref{p-m-c}), one obtains
\begin{eqnarray}
&&3(\lambda-1)\left(\dot\psi+H\phi+\frac{1}{3}\nabla^2\beta\right)+\eta\left(\dot\phi+H\phi\right)
\nonumber\\&&\quad
+\left(\dot{\delta{\cal H}}
+H\delta{\cal H}\right)-2H\delta{\cal E}+\frac{\delta{\cal F}}{\mpl^2}=0.\label{per_eq}
\end{eqnarray}

\subsection{Minkowski limit}

Let us check that the perturbation equations derived in the previous
subsection can reproduce the known result in the Minkowski background by
taking the limit $H\to 0$.  Since we are particularly interested
in the stability of a scalar graviton in the infrared (IR) regime, we
neglect the UV terms $\delta{\cal H}$ and $\delta{\cal E}$.  Then, the
Hamiltonian constraint simply gives $\phi = (2/\eta)\psi$.  The trace
part of the evolution equation~(\ref{p-ev-t}) reduces to
$\mathscr{A}\ddot\psi+\nabla^2\dot\beta=0,$ while the traceless
part~(\ref{p-ev-tl}) implies $\dot\beta=(1-2/\eta)\psi$.  We thus obtain
\begin{eqnarray}
\mathscr{A}\ddot\psi-\frac{2-\eta}{\eta}\nabla^2\psi=0.
\end{eqnarray}
This reproduces the result obtained in~\cite{BPS}.
It can be seen that the scalar graviton is stable in the IR regime
provided that
\begin{eqnarray}
0<\eta<2.  \label{cond_eta}
\end{eqnarray}
The propagation speed of the scalar graviton is given by
\begin{eqnarray}
c_g^2=\frac{2-\eta}{\eta}\frac{1}{\mathscr{A}}.
\end{eqnarray}

In addition to the theoretical bound~(\ref{ghostfree})
and~(\ref{cond_eta}), low energy phenomenology can put constraints on
the value of $\lambda$ and $\eta$.  A cosmological bound comes from the
fact that $G_c$ does not coincide with $G_N$.  The bound on the
difference is derived from the primordial abundance of He$^4$:
$|G_c/G_N-1|\lesssim 0.13$~\cite{BBN, BPS}, which in turn gives a mild
constraint on the parameters. This constraint can be further weakened by
considering additional relativistic degree of freedom and/or large
lepton asymmetry \cite{largeL} to compensate a change in the expansion
rate of the universe. More stringent constraints will come from the
parameterized post-Newtonian study.  See Ref.~\cite{BPS:comment} for a
preliminary discussion on this point.

\section{Evolution of cosmological perturbations in the IR}

In what follows, we are interested in perturbation modes whose spatial momenta
are much smaller than $\mpl$ and $M_{\rm m}$, so that
we ignore any UV terms in gravity and matter sectors.

\subsection{Superhorizon evolution}

We begin with examining the large scale evolution of perturbations analytically
by neglecting gradient terms.
It can be shown
from the matter equation of motion with $\delta{\cal F}=0$
that
\begin{eqnarray}
\zeta:=-\psi-H\frac{\delta\rho}{\dot\rho}
\end{eqnarray}
is conserved on large scales, $\zeta\simeq\zeta_*=$ constant,
in the case of a negligible non-adiabatic pressure perturbation.
This is the well-known fact when matter is conserved locally~\cite{LythWands}.
Using the Hamiltonian constraint, we obtain
\begin{eqnarray}
-\psi+\frac{H}{\dot H}\left(\dot\psi+H\phi\right)\simeq\zeta_*.\label{zeta**}
\end{eqnarray}
Equation~(\ref{per_eq}) on large scales reads
\begin{eqnarray}
\mathscr{B}\left(\dot\psi +H\phi\right)+\dot\phi+H\phi\simeq 0,
\end{eqnarray}
where $\mathscr{B}:=3(\lambda-1)/\eta\,(>0)$. The above two equations are combined to give
\begin{eqnarray}
\frac{\D^2\bar\psi}{\D\ln a^2}
-\left[\frac{(1-\epsilon)^2}{4}- \epsilon \mathscr{B}-
\frac{\D\epsilon}{\D\ln a}\right]\bar\psi\simeq 0,
\end{eqnarray}
where $\bar\psi:=(\psi+\zeta_*)\exp\left[\frac{1}{2}\int (1+\epsilon)\D\ln a\right]$
and $\epsilon :=-\D\ln H/\D\ln a$.
We may assume $\epsilon>0$ and $\D\epsilon/\D\ln a \simeq 0$.
(These assumptions are indeed true
in a matter-dominated universe and a radiation-dominated universe.)
Then, if $(1-\epsilon)^2-4\epsilon \mathscr{B}<0$, $\bar\psi$ oscillates
and therefore $\psi+\zeta_*$ decays as $\sim a^{-(1+\epsilon)}/2$.
If $(1-\epsilon)^2-4\epsilon \mathscr{B}>0$, $\bar\psi$ may grow like
$\bar\psi\sim a^p$ with $p=\sqrt{(1-\epsilon)^2-4\epsilon \mathscr{B}}/2$, but
$\psi+\zeta_*$ decays also in this case because $(1+\epsilon)/2>p$.
We therefore conclude that 
the large scale evolution of the non-decaying perturbation
can be characterized by a single constant $\zeta_*$, and
\begin{eqnarray}
\psi\simeq -\zeta_*,\quad \phi\simeq 0,\label{superhorizonsolution}
\end{eqnarray}
provided that the mode stays in the superhorizon regime for a sufficiently long time.
This indicates that {\em the subsequent evolution of cosmological perturbations
is essentially determined from the value of $\zeta_*$.}
This fact will be confirmed by a numerical calculation.

A remark is in order. In usual gravitational theories having general covariance,
$\zeta$ is conveniently computed from the matter sector by taking the gauge $\psi=0$.
In Ho\v{r}ava gravity, however, $\psi$ cannot be gauged away because
the symmetry is reduced and
the relevant gauge parameter is independent of $\Vec{x}$.
In this sense, two scalar-type contributions,
a scalar graviton $\psi$ and an adiabatic matter fluctuation $\delta\rho$,
translate to $\zeta$.
Therefore, we need to work out the two physically different contributions
in order to compute $\zeta$ in Ho\v{r}ava gravity.

\subsection{Subhorizon evolution}\label{sec:sub}

Let us define
\begin{eqnarray}
\delta:=\frac{\delta \rho+3H\delta J}{\rho}.
\end{eqnarray}
Using the equation of motion of matter, the momentum conservation equation,
the evolution equations, and the Hamiltonian constraint,
we find, for a universe dominated by a fluid with $w=p/\rho=$ const.,
\begin{eqnarray}
\ddot\delta+(2-3w)H\dot\delta &=&w\frac{\nabla^2}{a^2}\delta+
\frac{3}{2}(1-w)(1+3w)H^2\delta
\nonumber\\&&\;\;
+(1+w)\frac{3(\lambda-1)+\eta}{3\lambda-1}\frac{\nabla^2}{a^2}\phi,
\label{qs1}
\end{eqnarray}
which coincides with the corresponding equation in general relativity
except the last term.
The Hamiltonian and momentum constraints give
\begin{eqnarray}
2\frac{\nabla^2}{a^2}\psi-\eta\frac{\nabla^2}{a^2}\phi-2H\nabla^2\beta=\frac{\rho\delta}{\mpl^2},
\label{qs2}
\end{eqnarray}
which can be regarded as a generalized Poisson equation.
So far no approximation has been made other than neglecting the UV terms.

We now focus on the subhorizon evolution of $\delta$ in a
matter-dominated universe ($w=0$).  It is natural to assume that in the
matter-dominated universe the metric perturbations are slowly-varying
functions of time in the sense that their time derivatives give rise to
the Hubble scale: $\dot\psi, \dot\phi\sim H\psi,H\phi$.\footnote{Here we
dropped by hand the contribution of a scalar gravitational wave, but the
approximation will be justified by a numerical calculation.}
Under this ``quasi-static'' approximation, the evolution equations imply
that
\begin{eqnarray}
\psi\approx\phi,\quad\beta\approx 0,\label{subhorizonsolution1}
\label{qs3}
\end{eqnarray}
on subhorizon scales.
Equations~(\ref{qs1})--(\ref{qs3}) are then arranged to give
the following equations:
\begin{eqnarray}
&&\ddot\delta+2H\dot\delta = \frac{\nabla^2}{a^2}\phi,\label{dens1}
\\
&&\frac{\nabla^2}{a^2}\phi = 4\pi G_N\rho\,\delta,\label{dens2}
\end{eqnarray}
where $8\pi G_N=\mpl^{-2}(1-\eta/2)^{-1}$.
This gravitational constant
coincides with effective Newton's constant defined in~\cite{BPS}
(so that it differs from the gravitational constant in the Friedmann equation, $G_c$).
This is a natural extension of the result of~\cite{BPS},
where the gravitational field of a static point source has been derived.
Again, we have the relation $\psi=\phi$, in contrast to the case of Lorentz-invariant
scalar-tensor theories of gravity.

Equations~(\ref{dens1}) and~(\ref{dens2}) admit the analytic solution
\begin{eqnarray}
\delta = C_1t^{(-1+\sqrt{1+24\xi})/6}+C_2t^{(-1-\sqrt{1+24\xi})/6},\label{analytic}
\end{eqnarray}
where $\xi :=G_N/G_c=(3\lambda-1)/(2-\eta)$.
The first term grows in time, and,
for $\xi\neq 1$, the growth rate of the matter density perturbation
is slightly different from the standard one ($\delta\sim t^{2/3}$).

Next, let us consider a radiation-dominated universe ($w=1/3$) and
study the evolution of $\delta$ inside the sound horizon.
From Eq.~(\ref{qs2}) we may estimate $(\nabla^2/a^2)\phi\sim{\cal O}( H^2 \delta)$,
so that we have
$\ddot\delta+H\dot\delta\simeq (\nabla^2/3a^2)\delta$ inside the sound horizon.
This coincides with the corresponding equation in general relativity, and
the last term hinders the growth of the density perturbation in the radiation-dominated stage.
Moving to the Fourier space and using the conformal time defined by $\D \tau=\D t/a$,
this equation can easily be solved to give
$\delta = C_0\cos(k\tau/\sqrt{3}+\theta_0)$,
with $C_0$ and $\theta_0$ being integration constants.

Although the behavior of the density perturbation on subhorizon scales in the radiation-dominated universe
is simple as seen above, the metric perturbations evolve in a non-trivial manner.
Using the generalized Poisson equation~(\ref{qs2}),
the traceless part of the evolution equations~(\ref{p-ev-tl}), and Eq.~(\ref{per_eq}),
with some manipulation and the approximation $k^2\gg {\cal O}(1/\tau^2)$,
we arrive at
\begin{eqnarray}
\mathscr{A}\left(\phi''+
\frac{2}{\tau}\phi'+c_g^2k^2\phi\right)
\simeq 
-\frac{6}{\eta}\frac{a^2\rho}{\mpl^2}\left(\frac{\delta'}{k^2\tau}+\frac{\delta }{3}\right),
\end{eqnarray}
where ${}':=\partial_\tau$. The detailed derivation is
presented in Appendix~\ref{app1}.
A solution to this equation is given by
\begin{eqnarray}
\phi=
\Psi(\tau; k)+\frac{6k^{-2}}{\eta\mathscr{A}-3(2-\eta)}\frac{a^2\rho}{\mpl^2}
\left[\delta+{\cal O}(\delta'/k^2\tau)\right],\label{anlphi}
\end{eqnarray}
where $\Psi(\tau; k)$ is a solution to the source-free wave equation
$ \Psi''+ (2/\tau) \Psi'+c_g^2k^2\Psi =0$,
and hence
\begin{eqnarray}
\Psi = C_\Psi \frac{\cos\left(c_g k\tau+\theta_\Psi\right)}{a},\label{anlgws}
\end{eqnarray}
where $C_\Psi$ and $\theta_\Psi$ are integration constants.
The mode $\Psi$ can be naturally identified as the scalar graviton,
and it decays as $\sim a^{-1}$ inside the horizon, analogously to the usual
helicity-2 gravitational wave mode in a cosmological setting.
Note that the propagation speed of a scalar gravitational wave
in a cosmological background
is identical
to $c_g$ in the Minkowski background, and its stability is ensured for $c_g^2>0$.
The $\delta$ induced part of $\phi$ decays as $\sim a^{-2}$.
$\psi$ can be obtained from the generalized Poisson equation~(\ref{qs2}) as
\begin{eqnarray}
\psi\approx \frac{\eta}{2}\Psi
-\frac{1}{2k^2}\frac{\eta\mathscr{A}-3(2+\eta)}{\eta\mathscr{A}-3(2-\eta)}
\frac{a^2\rho}{\mpl^2} \delta.\label{anl_psi}
\end{eqnarray}
The subhorizon evolution in a radiation-dominated universe
is apparently characterized by four integration constants,
but they are completely determined by a single constant, $\zeta_*$,
and thus are in fact related to each other.


\subsection{Numerical solutions}

\begin{figure}[tb]
  \begin{center}
    \includegraphics[keepaspectratio=true,height=55mm]{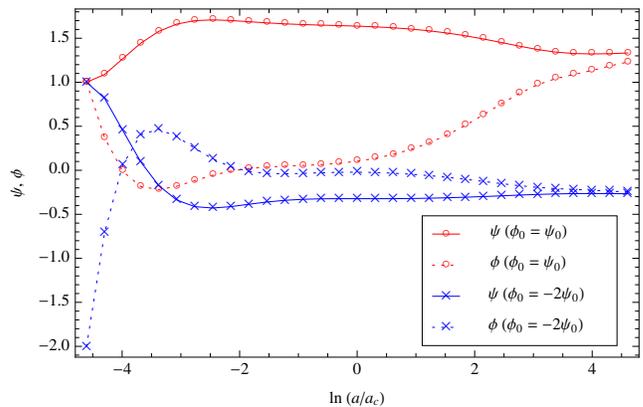}
  \end{center}
  \caption{Evolution of $\psi$ and $\phi$ in a matter-dominated universe.
  Red lines with circles correspond to the initial condition $\phi_0=\psi_0 (=1)$, while
  blue lines with crosses $\phi_0=-2\psi_0 (=-2)$. Other initial data are the same
  and are given by $\beta_0=1\times k^{-1}$ and $\chi_0=0$.
  The wavenumber is given by $k=0.1\times a(t_0)H(t_0)$.
  The parameters are $\lambda=1.05$ and $\eta=0.1$,
  and $a_c$ represents the ``horizon-crossing time'' defined by $k=a_cH(a_c)$.}%
  \label{fig:MD_psi.eps}
\end{figure}

\begin{figure}[tb]
  \begin{center}
    \includegraphics[keepaspectratio=true,height=55mm]{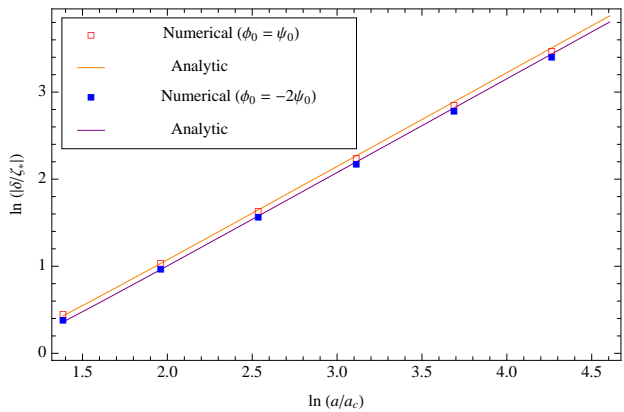}
  \end{center}
  \caption{Subhorizon evolution of the density perturbation (divided by $\zeta_*$)
  in a matter-dominated universe, compared to the analytic solution~(\ref{analytic}).}%
  \label{fig:MD_delta.eps}
\end{figure}

\begin{figure}[tb]
  \begin{center}
    \includegraphics[keepaspectratio=true,height=55mm]{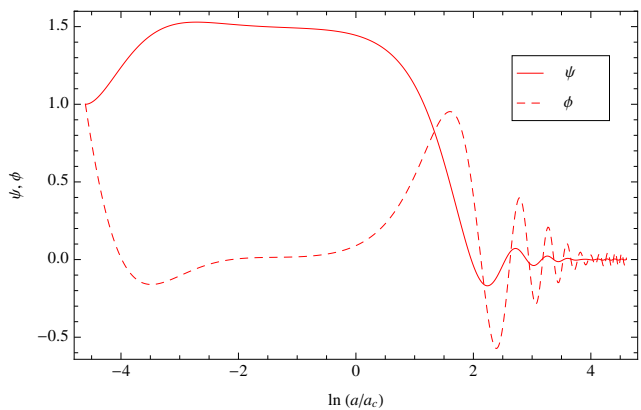}
  \end{center}
  \caption{Evolution of $\psi$ (solid line) and $\phi$ (dashed line)
  with $k=0.01\times a(t_0)H(t_0)$
  in a radiation-dominated universe.
  The initial condition is given by $\psi_0=1=\phi_0$, $\beta_0=0$, and $\chi_0=0$.
  The parameters are $\lambda=1.05$ and $\eta=0.1$, and $a_c$ is defined similarly to
  the previous figures: $k=a_cH(a_c)$.}%
  \label{fig:RD.eps}
\end{figure}

\begin{figure}[tb]
  \begin{center}
    \includegraphics[keepaspectratio=true,height=55mm]{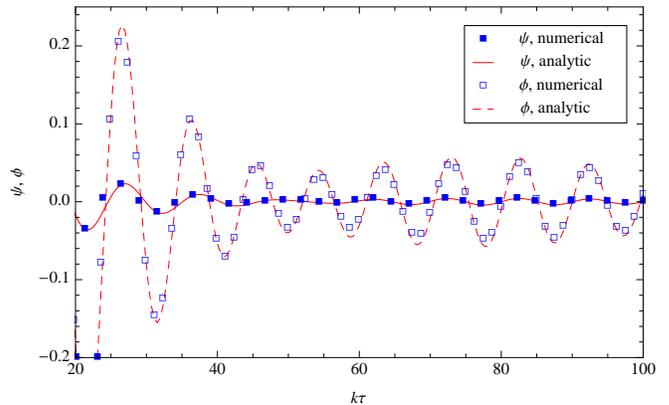}
  \end{center}
  \caption{Subhorizon evolution of $\psi$ and $\phi$ in a radiation dominated universe,
  compared to the analytic solutions~(\ref{anlphi})--(\ref{anl_psi}).
  The initial condition and parameters are the same as in Fig.~\ref{fig:RD.eps}.}%
  \label{fig:osc.eps}
\end{figure}

We have solved numerically
the perturbation equations in the IR without any other approximations.
The procedure for doing so is described in some detail in Appendix~\ref{app2}.
The results are summarized in Figs.~\ref{fig:MD_psi.eps}--\ref{fig:osc.eps},
all of which reproduce the analytic results obtained above.

In Fig.~\ref{fig:MD_psi.eps} we show the evolution of $\psi$ and $\phi$
in a matter-dominated universe.
The initial condition is set by specifying four numbers $\psi_0=\psi(t_0)$,
$\chi_0=\dot\psi(t_0)$, $\phi_0=\phi(t_0)$, and $\beta_0=\beta(t_0)$ at some
initial time $t=t_0$.
It can be seen that
the perturbation evolution starting with the two different initial conditions
first converges to the solution~(\ref{superhorizonsolution})
outside the horizon, and then to (\ref{subhorizonsolution1}) inside the horizon.
The evolution of $\delta$ is shown in Fig.~\ref{fig:MD_delta.eps} for
the same set of the initial conditions as in Fig.~\ref{fig:MD_psi.eps}.
One can confirm that the growth of $\delta$ is in agreement with the
analytic solution~(\ref{analytic}).

The same large scale behavior, $\psi\simeq-\zeta_*$ and $\phi\simeq 0$,
is found also in a radiation-dominated universe,
as can be seen in Fig.~\ref{fig:RD.eps}.
The oscillation on small scales in the radiation-dominated universe, shown in Fig.~\ref{fig:osc.eps},
is well approximated by the analytic solutions presented in the previous subsection.
The $\delta$ induced part and $\Psi$ typically have the same amplitude
in the plotted region, and $\Psi$ dominates for larger $\tau$.
We have confirmed that
different initial conditions with the same value of $\zeta_*$ (and the same set of the other parameters)
result in the identical profile.

\section{Conclusions}

In this paper, we have studied cosmological perturbations in a
healthfully extended version of Ho\v{r}ava gravity.  The most general
perturbation equations have been derived without specifying the matter
content.  We then solved the resultant perturbation equations in the IR
regime analytically and numerically, for a universe dominated by a
single fluid with vanishing non-adiabatic fluctuations.  It was found
that the large scale evolution converges to that described by a
constant $\zeta$, which is an analog of the curvature perturbation on
uniform density hypersurfaces commonly used in the context of usual
gravitational theories having general covariance.  This implies that,
although the system has two scalar degrees of freedom corresponding to a
scalar graviton and an adiabatic matter fluctuation, it is sufficient to
specify the value of $\zeta$ for predicting the late-time evolution of
perturbations.  Our analytic results were confirmed by numerical
calculations.

It would be important to revisit the analysis of cosmological
perturbations in extended Ho\v{r}ava gravity by employing the
Hamiltonian formulation in order to get a more transparent understanding
of the properties of the scalar graviton~\cite{K-A, GKS}.  This issue is
left for a further study.

It was pointed out very recently that the IR limit of extended
Ho\v{r}ava gravity is identical to Einstein-aether theory~\cite{EA} if
the aether vector is restricted to be hypersurface
orthogonal~\cite{BPS:comment, Jacobson}.  Hypersurface orthogonal
solutions of Einstein-aether theory are also solutions to the IR limit
of extended Ho\v{r}ava gravity, though the converse is not true.  This
observation is insightful for understanding the background cosmological
dynamics, as a cosmological background aether field is hypersurface
orthogonal. For example, the discrepancy between $G_N$ and $G_c$ is
a generic feature of Einstein-aether theory. Note, however, that the
aether field is no longer hypersurface orthogonal at perturbative order.
It would be interesting to compare the evolution of cosmological
perturbations in extended Ho\v{r}ava gravity with that in
Einstein-aether theory~\cite{EACP, Garriga}.

\acknowledgments 
We are grateful to Takeshi Chiba for useful comments. T.K. and
Y.U. are supported by the JSPS under Contact Nos.~19-4199 and
19-720. M.Y. is supported by JSPS Grant-in-Aid for Scientific research
No.\,21740187.

\appendix
\section{Master equations on small scales}\label{app1}

We present a detailed derivation of the master equations
governing the subhorizon evolution of $\delta$ and $\phi$
for general $w$ (=const.).

Defining convenient variables as $u:=a^2\psi$, $v:=a^2\phi$, and $\tilde\beta=a^3\beta$,
Eq.~(\ref{qs2}) can be written as
\begin{eqnarray}
2u-\eta v-2\frac{a'}{a}\tilde\beta
+\frac{a^4\rho}{\mpl^2}\frac{\delta}{k^2}= 0,\label{RD-sub1}
\end{eqnarray}
while the traceless part of the evolution equations reads
\begin{eqnarray}
\tilde\beta'+v-u= 0,\label{RD-sub2}
\end{eqnarray}
where the conformal time $\tau$ is used: ${}':=\partial_\tau$.
Equation~(\ref{per_eq}) implies
\begin{eqnarray}
\mathscr{B}\left(u'-2\frac{a'}{a}u+\frac{a'}{a}v
-\frac{k^2}{3}\tilde\beta\right)+v'-\frac{a'}{a}v= 0.\label{RD-sub3}
\end{eqnarray}
Differentiating Eq.~(\ref{RD-sub3}) and using Eq.~(\ref{RD-sub2}) to remove $\tilde\beta'$,
we obtain
\begin{eqnarray}
\mathscr{B}\left[u''-2\frac{a'}{a}u'+\frac{a'}{a}v'+\frac{k^2}{3}(v-u)\right]+v''-\frac{a'}{a}v'\simeq 0,
\nonumber\\ \label{apeq1}
\end{eqnarray}
where we used $a''/a, (a'/a)^2\ll k^2$. The expression in the square brackets
can be written, using Eq.~(\ref{RD-sub1}) and then Eq.~(\ref{RD-sub2}), as
\begin{eqnarray}
\frac{\eta}{2}\left(v''-2\frac{a'}{a}v'\right)+\frac{2-\eta}{2}\frac{k^2}{3}v
-\frac{a^4\rho}{2\mpl^2}{\cal M}
\nonumber\\
+\frac{a'}{a}\left(u'-\frac{k^2}{3}\tilde \beta\right)=0,\label{apeq2}
\end{eqnarray}
where
\begin{eqnarray}
{\cal M}:=\frac{\delta''}{k^2}-6w\frac{a'}{a}\frac{\delta'}{k^2}-\frac{\delta}{3},
\end{eqnarray}
and
we used $k^2\gg{\cal O}(\tau^{-2})$ again.
Using Eq.~(\ref{RD-sub3}), the expression in the second line can be evaluated as
$\simeq -\mathscr{B}^{-1}(a'/a)v'$.
Thus, we arrive at
\begin{eqnarray}
\mathscr{A}\left(v''-2\frac{a'}{a}v'+c_g^2k^2v\right) = \frac{3}{\eta}\frac{a^4\rho}{\mpl^2}{\cal M},
\end{eqnarray}
or equivalently,
\begin{eqnarray}
\mathscr{A}\left(\phi''+2\frac{a'}{a}\phi'+c_g^2k^2\phi\right) =
\frac{3}{\eta}\frac{a^2\rho}{\mpl^2}{\cal M}.\label{phiequation}
\end{eqnarray}

The evolution equation of $\delta$, Eq.~(\ref{qs1}), can be written in terms of the conformal time as
\begin{eqnarray}
&&\delta''+(1-3w)\frac{a'}{a}\delta'+wk^2\delta=\frac{3}{2}(1-w)(1+3w)
\left(\frac{a'}{a}\right)^2\delta
\nonumber\\&&\qquad\qquad\qquad\qquad\qquad
-(1+w)\frac{3(\lambda-1)+\eta}{3\lambda-1}k^2\phi.\label{deltaequation}
\end{eqnarray}
We should note that here we do not drop the ${\cal O}\left((a'/a)^2\delta\right)$ term
because ${\cal O}(k^2\delta)$ term vanishes for $w=0$, in contrast to
the approximation made in deriving Eqs.~(\ref{apeq1}), (\ref{apeq2}), and (\ref{phiequation}),
where we implicitly assumed that the coefficients of ${\cal O}(k^2)$ terms
are not too far from the order of unity value.
In particular, we are not considering the case with $c_g^2\ll 1$.
Using Eq.~(\ref{deltaequation}), we have
\begin{eqnarray}
{\cal M}=-(1+3w)\left(\frac{a'}{a}\frac{\delta'}{k^2}+\frac{\delta}{3}\right)+{\cal O}(\phi).
\end{eqnarray}
Note that $\delta\sim(k/aH)^2\phi\gg\phi$ on small scales.

In a matter-dominated universe, $w=0$, the following
solves Eqs.~(\ref{phiequation}) and~(\ref{deltaequation}):
\begin{eqnarray}
\delta& =& C_1 a^{(-1+\sqrt{1+24\xi})/4}+C_2 a^{(-1-\sqrt{1+24\xi})/4},\label{appsol1}
\\
\phi&=&-\frac{a^2}{k^2}\cdot 4\pi G_N\rho\,\delta+\Psi,\label{appsol2}
\end{eqnarray}
where $\xi$ and $G_N$ are defined in the main text, and
$\Psi$ is a solution to the wave equation $\Psi''+2(a'/a)\Psi'+c_g^2k^2\Psi=0$,
and hence can be identified as a scalar gravitational wave.
$\Psi$ decays as $\sim a^{-1}$ on small scales.
We assumed that the amplitude of $\Psi$ is small enough not to source $\delta$ via Eq.~(\ref{deltaequation}).
Neglecting $\Psi$, Eqs.~(\ref{appsol1}) and~(\ref{appsol2}) reproduce the result obtained in the main text.

\section{Solving the perturbation equations numerically}\label{app2}

In this appendix we describe the procedure for solving the perturbation equations numerically.
The UV terms $\delta{\cal H}$, $\delta{\cal E}$, and $\delta{\cal F}$ are all neglected.
We use the following variables:
$\chi:=\dot\psi$, ${\cal D} :=\delta\rho /(\rho+p)$, and ${\cal J} :=\delta J /(\rho+p)$.
Initial data at $t=t_0$ are specified as follows:
\begin{eqnarray}
\psi(t_0)=\psi_0,\;\;
\chi(t_0)=\chi_0,\;\;
\phi(t_0) = \phi_0,
\;\; \beta(t_0)=\beta_0.
\end{eqnarray}
Using the constraint equations, we then
obtain the initial conditions for ${\cal D}$ and ${\cal J}$ at $t=t_0$.
Choosing $t_0$ in the superhorizon regime, we
can evaluate $\zeta_*$ from Eq.~(\ref{zeta**}).

The constraint equations can be solved for $\phi$ and $\beta$ (in the Fourier space) as follows:
\begin{eqnarray}
\mathscr{L}\phi&=&-\mathscr{A}\left[2\frac{\chi}{H}+(\lambda-1)\frac{\dot H}{H^2}
\left({\cal D}+\mathscr{A}H {\cal J}\right)\right]
\nonumber\\&&\quad +2\frac{k^2}{a^2H^2}\psi,
\\
\mathscr{A}^{-1}\mathscr{L}k^2\beta&=&\frac{k^2}{a^2H^2}\left(\eta \chi
+2H\psi+\eta\dot H{\cal J}\right)
\nonumber\\&&\quad
-(3\lambda-1)\frac{\dot H}{H}\left({\cal D}+3H{\cal J}\right),
\end{eqnarray}
where $\mathscr{L}:=\eta(k/aH)^2+2\mathscr{A}$.
These two equations can be used to remove $\phi$ and $\beta$ in the
other equations, so that a system of four first-order equations for
$\psi$, $\chi$, ${\cal D}$, and ${\cal J}$ can be derived.
Assuming $\delta p-c_s^2\delta\rho=0$ with $c_s^2:=\dot p/\dot\rho$,
we obtain
\begin{eqnarray}
\dot{{\cal D}}&=&3\chi-\frac{k^2}{a^2}{\cal J}-k^2\beta,
\\
\dot{{\cal J}}&=&3Hc_s^2{\cal J}+\phi+c_s^2{\cal D},
\end{eqnarray}
from the matter equations, and
\begin{eqnarray}
&&\mathscr{A}\left[\dot \chi+3H\chi+H\dot\phi +\left(3H^2+2\dot H\right)\phi\right]
\nonumber\\&&\quad
-\frac{k^2}{a^2}
(\psi  -\phi )=-\mathscr{A}\dot Hc_s^2{\cal D},
\end{eqnarray}
from the evolution equations,
together with $\dot\psi = \chi$.
In the main text the above set of the equations is solved for $c_s^2=w=$ constant.


\end{document}